## ORIGINAL ARTICLE

# Dynamic Modeling of the Interaction Between Autophagy and Apoptosis in Mammalian Cells

I Tavassoly[1,†], J Parmar[2,†], AN Shajahan-Haq[3], R Clarke[3], WT Baumann[4] and JJ Tyson[2,*]

**Autophagy is a conserved biological stress response in mammalian cells that is responsible for clearing damaged proteins and organelles from the cytoplasm and recycling their contents via the lysosomal pathway. In cases of mild stress, autophagy acts as a survival mechanism, while in cases of severe stress cells may switch to programmed cell death. Understanding the decision process that moves a cell from autophagy to apoptosis is important since abnormal regulation of autophagy occurs in many diseases, including cancer. To integrate existing knowledge about this decision process into a rigorous, analytical framework, we built a mathematical model of cell fate decisions mediated by autophagy. Our dynamical model is consistent with existing quantitative measurements of autophagy and apoptosis in rat kidney proximal tubular cells responding to cisplatin-induced stress.**

Autophagy and its dysregulation play important roles in the pathogenesis of many complex diseases.[1] For instance, autophagy helps cancer cells to survive the stresses of nutrient deprivation and anoxia.[2] Autophagy is also involved in the development of resistance to chemotherapy; inhibiting autophagy can increase the therapeutic responses of resistant cancer cells to chemotherapy, endocrine therapy, or radiation therapy.[3,4] While autophagy is normally initiated as a prosurvival response to stress, excessive stress can trigger cell death.

Recently, we proposed a systems biology approach to model the complex interplay among pathways for estrogen and growth factor signaling, unfolded protein response (UPR) stress, autophagy, and apoptosis in the context of breast cancer responses to endocrine therapy.[5] Other authors as well have argued that mathematical theories of the systems-level properties of molecular signaling networks will play pivotal roles in the emerging field of systems pharmacology.[6] In particular, several mathematical models of autophagy have been proposed recently. Martin *et al.*[7] presented a computational model of autophagic vesicle dynamics in single cells, but they did not address the crucial interplay between autophagy and apoptosis. Kapuy *et al.*[8] addressed this interplay using a simplified mathematical model, but they did not compare their simulations with experimental measurements of how cells respond to stress. Our motivation for building mathematical models of autophagy and apoptosis is to integrate current mechanistic knowledge of these processes into a coherent framework and to determine if the mechanisms we include can explain existing qualitative observations and quantitative data on autophagic responses of cells to stress. As more data become available, this model can serve as a foundation for better models, with the goal of accurately predicting how therapeutic interventions may alter cell fates in normal and diseased tissues.

## MOLECULAR CELL BIOLOGY OF AUTOPHAGY AND APOPTOSIS

Autophagy is a conserved catabolic cellular process by which a cell degrades its own components, including damaged proteins and organelles. Autophagosomes (subcellular organelles enclosed by two or more membranes) engulf damaged materials, fuse with lysosomes, and the resulting autolysosome then uses lysosomal enzymes to degrade the contents of the autophagic vacuoles.[1] Autophagy-related proteins (ATG proteins) drive autophagosome formation in yeast and mammalian cells.[1,9] Commitment of a cell to autophagy seems to occur at the earliest stages of vesicle nucleation and formation of the isolation membrane, a small, flattened membrane sac that elongates and curves to create an autophagosome.[1,9]

Major molecular players in the induction of autophagy in mammalian cells are mTOR (the mammalian target of rapamycin) and ATG13. mTOR is a signal integrator that senses stress conditions such as endoplasmic reticulum (ER) stress, hypoxia, low growth factor levels, or low levels of essential amino acids.[10] When there are no critical stress conditions in the cell, a protein complex consisting of mammalian ATG13, Unc-51-like autophagy activating kinase 1 (ULK1, the mammalian homolog of yeast Atg1), and focal adhesion kinase interacting protein of 200 kD (FIP200) is repressed by mTOR phosphorylation of ATG13 and ULK1. Cellular stress inactivates mTOR, allowing the

[1]Graduate Program in Genetics, Bioinformatics and Computational Biology, Virginia Polytechnic Institute and State University, Blacksburg, Virginia, USA; [2]Department of Biological Sciences, Virginia Polytechnic Institute and State University, Blacksburg, Virginia, USA; [3]Lombardi Comprehensive Cancer Center, Georgetown University Medical Center, Washington, DC, USA; [4]Bradley Department of Electrical and Computer Engineering, Virginia Polytechnic Institute and State University, Blacksburg, Virginia, USA. Correspondence: JJ Tyson (tyson@vt.edu)
[†]These authors contributed equally to this work.
Present affiliation for I Tavassoly: Department of Pharmacology & Systems Therapeutics, Mount Sinai School of Medicine, New York, NY, USA.

ULK1:ATG13:FIP200 complex to be active.[11] The active complex promotes formation of the isolation membrane.

Beclin-1, the mammalian ortholog of yeast Atg6, is necessary for autophagosome formation, playing a key role in vesicle nucleation.[1,9] BCL-2 family proteins in the ER function as antiautophagy proteins through their inhibitory binding with Beclin-1. Although Beclin-1 contains a BCL-2 homology domain 3 (BH3), it is not proapoptotic.[1,12] For autophagosome formation to begin, Beclin-1 must be released from BCL-2 inhibition, which is promoted by either phosphorylation of BCL-2 by c-Jun N-terminal kinase (JNK), or by phosphorylation of Beclin-1 by death-associated protein kinase (DAPK).[13–16] Free Beclin-1 then binds with other partners to form a "Beclin-1 core complex," which promotes vesicle nucleation.[1,9]

For more details about the physiology and molecular biology of autophagy, we refer the reader to our review article.[3]

A convenient quantitative measure of autophagosome formation in mammalian cells is the state of microtubule-associated protein light chain 3 (LC3),[17] which is a mammalian homolog of yeast Atg8. LC3 exists in two forms: LC3-I (cytosolic form, 18 kDa) and LC3-II (membrane-bound form, 16 kDa). After autophagy initiation, LC3-I is converted to LC3-II, which then participates in the vesicle elongation step of autophagosome formation.[1,17] LC3-II can be distinguished from LC3-I by immunoblotting. Alternatively, autophagosome formation can be visualized as green "puncta" by tagging LC3 with green fluorescent protein.[18]

In contrast to autophagy, the molecular regulatory system that controls apoptosis is reasonably well understood. The extrinsic and intrinsic signaling pathways leading to caspase-dependent apoptosis have been studied by mathematical modeling.[5,19,20] We focus on the intrinsic pathway, which leads to mitochondrial outer membrane permeabilization (MOMP), causing release of cytochrome C into the cytoplasm and subsequent activation of executioner caspases. The induction of MOMP is dependent on oligomerization of proapoptotic proteins (BAX or BAK) in the outer mitochondrial membrane, while antiapoptotic BCL-2 family proteins in the mitochondria inhibit these proapoptotic proteins. Activation of BAX and BAK is triggered by BH3-only proteins.[1,19]

### Crosstalk between autophagy and apoptosis

BCL-2 family proteins in the ER and mitochondria are important regulators of autophagy and apoptosis, respectively.[21,22] Hence, different levels of expression of BCL-2 proteins in the ER membrane and in the mitochondrial membrane may lead to different activation dynamics for autophagy and apoptosis. Calcium signaling from the ER to mitochondria may also play a role in autophagy-apoptosis crosstalk.[21–25] The inositol-1,4,5-trisphosphate receptor (IP3R) and the sarco/endoplasmic reticulum $Ca^{2+}$-ATPase (SERCA) pump are the central regulators of $Ca^{2+}$ exchange between ER and cytoplasm. By pumping $Ca^{2+}$ from the cytosol into the ER, SERCA is responsible for maintaining very low calcium ion concentrations in the cytoplasm. Conversely, IP3R is a stress-activated $Ca^{2+}$ channel that releases $Ca^{2+}$ from the ER into the cytoplasm.[22–26]

Normally, IP3R is sequestered by BCL-2 family proteins in the ER membrane.[27] Phosphorylation of BCL-2 proteins dissociates the complex and allows for calcium release from the ER.[28] Sustained, elevated cytoplasmic $Ca^{2+}$ can lead to apoptosis.[22,23,29] Cytoplasmic $Ca^{2+}$ can also inhibit mTOR via activation of calmodulin-dependent kinase kinase-$\beta$ (CaMKK$\beta$), which activates AMPK (5′ AMP-activated protein kinase). AMPK has an inhibitory effect on mTOR.[30]

Calcium influx into mitochondria can induce apoptosis directly, and several other signaling pathways also link sustained calcium elevation to apoptosis.[22,23,29,31,32] For example, calcium activates calcineurin, which dephosphorylates and activates BAD, a proapoptotic BCL-2 family protein capable of inducing apoptosis.[33] In addition, calcium can activate calpain, a cysteine protease that cleaves ATG5, an essential protein for autophagosome formation. Truncated ATG5 induces apoptosis by suppressing antiapoptotic BCL-2 proteins in the mitochondria.[1,34,35] The mechanism of this suppression is unknown; so, for modeling purposes, we assume that truncated ATG5 upregulates proapoptotic BH3 proteins. During apoptosis, the activation of caspase 8 downregulates autophagosome formation by cleaving Beclin-1.[36–38]

For more details about the interplay of autophagy and apoptosis, we refer the reader to the excellent review article by Marino *et al*.[39]

### METHODS: MATHEMATICAL MODELING APPROACH

From these facts we draw an "influence diagram" (**Figure 1**), which summarizes our hypothesis about the most relevant molecular interactions between autophagy and apoptosis in mammalian cells. In **Supplementary Text S1**, we convert this influence diagram into a mathematical model. Our mathematical model is formulated in terms of 21 variables (**Table 1**) whose response to stress is described by the differential and algebraic equations in **Table 2**. The equations were solved numerically using the MatLab code (MathWorks, Natick, MA) in **Supplementary Text S2**. As described in **Supplementary Text S3**, we fitted model simulations to experimental data (**Supplementary Table S1**) to obtain the "optimal" set of parameter values in **Table 3**. Using these optimal parameter values, we computed the steady-state values of all variables under no-stress conditions ($C = 0$; see **Table 1**). These values were used as initial conditions for simulations of how cells respond to stress ($C > 0$).

To examine the generic properties of our model, we solve the governing equations for fixed (optimal) values of the parameters and varying levels of stress, $C$. These simulations represent how an "average" cell might respond to stress, but they cannot be compared to the observed behavior of a population of cells responding to cisplatin, because, in the latter case, we must take into account the heterogeneous response of cells to apoptotic signals. We attribute this heterogeneity to differences among cells in the mitochondrial concentration of BCL2, because (in our model) this parameter most sensitively determines whether or not apoptosis occurs, and if so, when. Using randomly selected [BCL2$_{mit}$] levels from a lognormal distribution, we simulate 100 cells and record the average value of every variable, including the percentage of cells having undergone apoptosis at each timepoint.

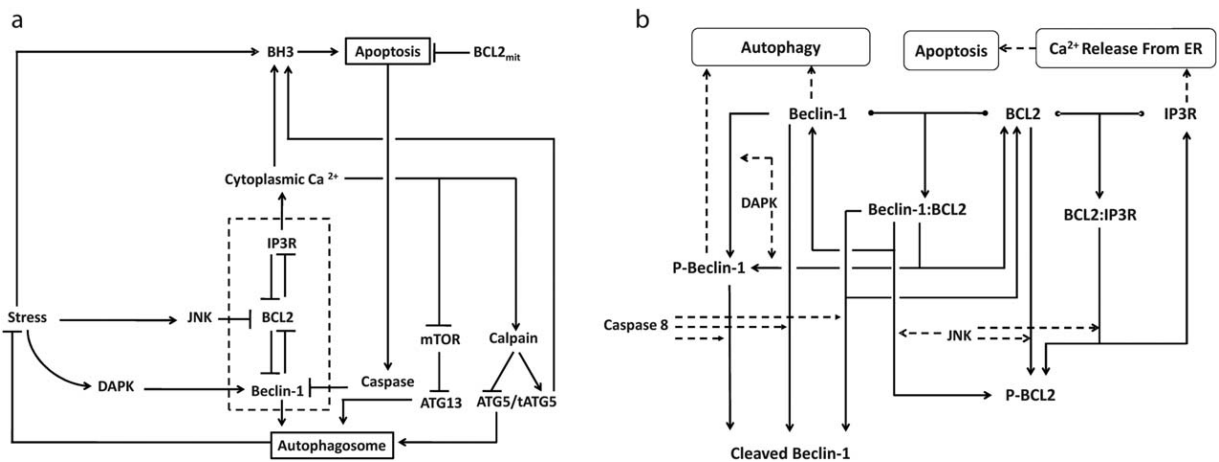

**Figure 1** The interplay between autophagy and apoptosis. (**a**) Diagram of the "influences" (activation = barbed arrows, inhibition = blunt arrows) between the major proteins controlling autophagosome formation and apoptosis. (**b**) More detailed diagram of the reactions involved in the dashed box in panel a. Solid arrows = chemical reactions; dashed arrows = catalytic activities; T-junctions = reversible formation of a binary complex.

Although we attribute cellular heterogeneity entirely to fluctuations in [BCL2$_{mit}$], this assumption is clearly an oversimplification. Other sources of variability, e.g., in BH3 production or ATG5 cleavage, may well contribute to the variable commitment of cells to apoptosis. Nonetheless, our assumption is a simple and effective way to fit the model to experimental observations of percent apoptosis.

Before we can compare our model to data (which has some level of uncertainty), we must quantify how uncertain we are about the optimal parameter values given in **Table 3**. To this end, we describe, in **Supplementary Text S3**, how we perturbed the optimal set of parameter values to obtain alternative sets of parameter values that provide "acceptable" fits of model simulations to the experimental data. We created a collection of 3,758 "acceptable sets of parameter values" and all simulations of experimental data are based on samples from this collection. In this way we take into account our uncertainty about the parameter values and the consequent range of predictions that are made by the model. In **Table 3** we record the coefficient of

**Table 1** Variables, their descriptions, and their values when cisplatin = 0

| Variable | Description | Steady-state value (no drug treatment) |
|---|---|---|
| [ATG5] | Concentration of active ATG5 | 0.717 |
| [tATG5] | Concentration of truncated ATG5 | 0.283 |
| [ATG13] | Concentration of active ATG13 protein | 0.0184 |
| [ATPHG] | Concentration of autophagosomes in cytoplasm | 0.285 |
| [BCL2]$_U$ | Concentration of unphosphorylated BCL-2 family proteins in ER | 2.463 |
| [BCL2_P] | Concentration of phosphorylated BCL-2 family proteins in ER | 0.537 |
| [BECN1]$_T$ | Concentration of total Beclin-1 protein | 3 |
| [BECN1_P] | Concentration of phosphorylated Beclin-1 protein | 0.0382 |
| [BECN1]$_F$ | Concentration of Beclin-1 protein free from suppression by BCL-2 | 1.121 |
| [BECN1]$_U$ | Concentration of unphosphorylated form of Beclin-1 protein | 2.962 |
| [BH3] | Concentration of active BH3 proteins | 0.0690 |
| [Ca$^{2+}$] | Concentration of cytoplasmic Ca$^{2+}$ | 0.397 |
| [CALPAIN] | Concentration of active CALPAIN | 0.0221 |
| [CASP] | Concentration of active caspase | 0 |
| [DAPK] | Concentration of active death-associated protein kinase | 0.103 |
| [IP3R]$_F$ | Concentration of IP3 receptors free from suppression by BCL-2 | 0.378 |
| [JNK] | Concentration of active c-Jun N-terminal kinase | 0.281 |
| [LIG]$_T$ | Concentration of total ligands available for binding to BCL-2 in ER | 3.962 |
| [LIG]$_F$ | Concentration of ligands free from suppression by BCL-2 | 1.499 |
| [MTOR] | Concentration of active mammalian target of rapamycin (mTOR) | 0.335 |
| S | Level of stress induced in the cell by drug treatment or other stressors | 0.831 |

**Table 2** Equations defining the model

**Differential Equations:**

$\frac{d[ATG5]}{dt} = \gamma_A \cdot (F(\sigma_A W_A) - [ATG5])$

$\frac{d[ATG13]}{dt} = \gamma_G \cdot (F(\sigma_G W_G) - [ATG13])$

$\frac{d[ATPHG]}{dt} = k_a \cdot ([BECN1]_F + [BECN1\_P])[ATG13][ATG5] - k_{da} \cdot [ATPHG]$

$\frac{d[BCL2\_P]}{dt} = \gamma_B \cdot ([BCL2_{ER}] \cdot F(\sigma_B W_B) - [BCL2\_P])$

$\frac{d[BECN1]_T}{dt} = -k_{casp} \cdot [CASP] \cdot [BECN1]_T$

$\frac{d[BECN1\_P]}{dt} = \gamma_L \cdot ([BECN1]_T \cdot F(\sigma_L W_L) - [BECN1\_P])$

$\frac{d[BH3]}{dt} = \gamma_H \cdot (F(\sigma_H W_H) - [BH3])$

$\frac{d[Ca^{2+}]}{dt} = k_{out} \cdot [IP3R]_F \cdot ([Ca^{2+}]_T - [Ca^{2+}]) - k_{in} \cdot [Ca^{2+}]$

$\frac{d[CALPAIN]}{dt} = \gamma_C \cdot (F(\sigma_C W_C) - [CALPAIN])$

$\frac{d[DAPK]}{dt} = \gamma_D \cdot (F(\sigma_D W_D) - [DAPK])$

$\frac{d[JNK]}{dt} = \gamma_J \cdot (F(\sigma_J W_J) - [JNK])$

$\frac{d[MTOR]}{dt} = \gamma_T \cdot (F(\sigma_T W_T) - [MTOR])$

$\frac{dS}{dt} = k_c \cdot C + k_{sb} - (k_{rb} + k_{ra} \cdot [ATPHG]) \cdot S$

**Algebraic Equations:**

$[tATG5] = 1 - [ATG5]$

$[BCL2]_U = [BCL2_{ER}] - [BCL2\_P]$

$[BECN1]_U = [BECN1]_T - [BECN1\_P]$

$[LIG]_T = [IP3R]_T + [BECN1]_U$

$[LIG]_F = \max(0, [LIG]_T - [BCL2]_U)$

$[IP3R]_F = [IP3R]_T \cdot \frac{[LIG]_F}{[LIG]_T}$

$[BECN1]_F = [BECN1]_U \cdot \frac{[LIG]_F}{[LIG]_T}$

$[CASP](0) = 0;\quad [CASP](t) = 1$ when $[BH3] = [BCL2_{mit}]$ and ever thereafter

**Definitions:**

$F(x) = \frac{1}{1+e^{-x}}$

$W_A = w_{ATG5\_0} - [CALPAIN]$

$W_B = -w_{BCL2P\_0} + [JNK]$

$W_C = -w_{CALP\_0} + [Ca^{2+}]$

$W_D = -w_{DAPK\_0} + S$

$W_G = w_{ATG13\_0} - [MTOR]$

$W_H = -w_{BH3\_0} + w_{BH3\_S} \cdot S + w_{BH3\_Ca} \cdot [Ca^{2+}] + [tATG5]$

$W_J = -w_{JNK\_0} + S$

$W_L = -w_{BCEN1P\_0} + [DAPK]$

$W_T = w_{MTOR\_0} - [Ca^{2+}]$

variation (CV = standard deviation/|mean|) of each parameter value over the collection of acceptable parameter sets. These CV's fall in the range 8%–19%.

## RESULTS

The experimental data we seek to explain involve the response of rat kidney proximal tubule (RPT) cells to treatment with cisplatin, a widely and effectively used antineoplastic drug. Cisplatin induces the UPR and activation of JNK and DAPK in mammalian cells.[40–42] Periyasamy-Thandavan *et al.*[18] used RPT cells, transiently transfected with GFP-LC3 and treated with cisplatin, to investigate the cytoprotective role of autophagy. To study the time course of autophagosome formation, the investigators measured temporal changes in cisplatin-induced LC3 puncta by fluorescence microscopy, and LC3-II production by immunoblotting. They also measured the autophagic response and the percentage of apoptotic cells when RPT cells were treated with cisplatin in the presence or absence of bafilomycin (BAF) and 3-methyladenine (3-MA), which are inhibitors of autophagy. To compare our model calculations with experimental data, we equated the autophagy level in our model, [ATPHG], with the LC3-II level measured by immunoblotting. Programmed cell death in our model is associated with an "indicator function" [CASP], which takes the value 0 in living cells and the value 1 when a cell commits to apoptosis, which occurs as soon as [BH3] exceeds [BCL2$_{mit}$].

### Generic properties of the model

The behavior of an "average" cell to cisplatin-induced stress is predicted by simulating the equations in **Table 2** with the optimal parameter values in **Table 3** (with [BCL2$_{mit}$] fixed at 0.12), starting from the steady-state initial conditions in **Table 1**. In **Figure 2a,b** we show how 13 of our variables change over the course of time for an average cell responding to a continuous dose of cisplatin, $C = 20$ in the model. (For time courses of all variables in the model, see **Supplementary Figure S1**.) Most changes occur within the first few hours, as the cells quickly activate autophagy to respond to the stress. Then, at $t = 18$ hours, executioner caspases are activated and the autophagic response switches off (**Figure 2c**) as the cell becomes apoptotic. These general characteristics of the response are consistent with the behavior of cells exposed to 20 μM cisplatin.[18] Indeed, parameter values in the model are chosen so that simulation results at $C = 20$ correspond to experimental results at a cisplatin dose of 20 μM.

In **Figure 2c** we show how the level of autophagy, over the course of 4 days, responds to a range of cisplatin doses up to $C = 100$. For low doses, [ATPHG] rises quickly (6–8 hours) to a steady-state level, in order to counteract the stress caused by cisplatin. For larger doses, [ATPHG] rises quickly but then falls to zero as BH3 proteins accumulate in the overstressed cells, which commit apoptosis when [BH3] = [BCL2$_{mit}$] = 0.12. In **Figure 2c**, one can see the activation of apoptosis as a kink in [ATPHG]($t$) when [CASP] switches from 0 to 1. For $C = 100$, 20, and 6, CASP is activated at $t = 7.5$, 18, and 59 hours, respectively. The apoptotic response (in our model) is driven primarily by calpain-dependent ATG5 cleavage, which dials back the production of autophagosomes and produces proapoptotic, truncated ATG5 molecules. Presumably these effects are intended to force cells to commit to apoptosis under conditions of high stress. Caspase activation cleaves Beclin-1 and turns off formation of autophagosomes. Beyond this timepoint, the remaining autophagosomes fuse with lysosomes and are degraded, with [ATPHG] ultimately decaying to zero.

In **Figure 2d** we repeat this simulation for a population of 100 cells with [BCL2$_{mit}$] following a lognormal distribution and plot the population-average level of autophagy at $t = 100$ hours. (We simulate to $t = 100$ hours to be reasonably sure that the dynamical model has reached its steady-state response.) For $C < 5$, autophagy ramps up with stress level and most cells survive. For $5 < C < 6$, the average level of autophagy drops with increasing $C$ because some—but not all—cells in the population die. For $C > 6$, all cells in the population are dead by $t = 100$ hours. To explore this switch between autophagy and apoptosis more closely, we plot (in **Figure 2e**) the percent apoptosis as a function of cisplatin dose at various timepoints from 6 hours to 100 hours.

Table 3 Parameters, their descriptions, their optimal values, and their coefficients of variation over the collection of acceptable parameter sets

| Parameters | Description | Optimal values[a] | Coeff. var.[a] |
|---|---|---|---|
| $C$ | Function of cisplatin dose | 0 | |
| $k_a$, $k_{da}$ | Rate constants for autophagosome formation and degradation (h$^{-1}$) | 1.77, 0.0948 | 15%, 14% |
| $k_c$ | Drug-induced stress rate ($\mu$M$^{-1}$ h$^{-1}$) | 0.51 | 14% |
| $k_{casp}$ | Rate constant for Beclin-1 cleavage by Caspase (h$^{-1}$) | 2.01 | 16% |
| $k_{in}$, $k_{out}$ | Rate constants for Ca$^{2+}$ transport from ER to cytoplasm and vice versa (h$^{-1}$) | 9.64, 6.31 | 14%, 14% |
| $k_{ra}$ | Rate constant for autophagic relief of stress (h$^{-1}$) | 3.83 | 18% |
| $k_{rb}$ | Rate constant for background relief of stress (h$^{-1}$) | 1.41 | 13% |
| $k_{sb}$ | Basal rate of stress (h$^{-1}$) | 2.08 | 13% |
| $w_{ATG5\_0}$, $w_{ATG13\_0}$ | Offsets of sigmoidal function when there are no signals | 0.215, 0.144 | 15%, 11% |
| $w_{BCL2P\_0}$, $w_{BECN1P\_0}$ | | 0.614, 0.647 | 11%, 10% |
| $w_{BH3\_0}$, $w_{CALP\_0}$ | | 1.26, 0.514 | 8%, 10% |
| $w_{DAPK\_0}$, $w_{JNK\_0}$ | | 2.98, 1.22 | 15%, 12% |
| $w_{MTOR\_0}$ | | 0.202 | 15% |
| $w_{BH3\_S}$, $w_{BH3\_Ca}$ | Interaction coefficients | 0.08, 0.0003 | 18%, 15% |
| $\gamma_A$, $\gamma_B$ | Rate constants for changes in protein concentrations (h$^{-1}$) | 0.524, 5.21 | 13%, 14% |
| $\gamma_C$, $\gamma_D$ | | 1.95, 4.05 | 15%, 13% |
| $\gamma_G$, $\gamma_H$ | | 1.04, 0.01 | 17%, 19% |
| $\gamma_J$, $\gamma_L$ | | 1.73, 3.43 | 11%, 16% |
| $\gamma_T$ | | 1.61 | 13% |
| $\sigma_A$, $\sigma_B$ | Steepness of sigmoidal response curves | 4.83, 4.57 | 14%, 11% |
| $\sigma_C$, $\sigma_D$ | | 32.3, 1.01 | 14%, 15% |
| $\sigma_G$, $\sigma_H$ | | 20.8, 2.89 | 12%, 12% |
| $\sigma_J$, $\sigma_L$ | | 2.42, 7.99 | 16%, 17% |
| $\sigma_T$ | | 3.51 | 12% |
| [BCL2$_{ER}$] | Total BCL-2 family proteins in ER | 3 | fixed |
| [BCL2$_{mit}$] | Total antiapoptotic BCL-2 family proteins in mitochondria | Mean = 0.120 | fixed |
| | (lognormally distributed) | SD = 0.0292 | fixed |
| [Ca$^{2+}$]$_T$ | Maximum cytoplasmic [Ca$^{2+}$] due to release of ER calcium | 2 | fixed |
| [IP3R]$_T$ | Total IP3R proteins in ER | 1 | fixed |

[a]$C(\mathbf{p}_{opt})$ = 0.5364, where $C(\mathbf{p})$ = cost function defined in Suppl. Text S2, for a parameter vector $\mathbf{p}$.
[b]Coefficient of variation = (SD) / |mean|. Note: the mean value of a parameter $\neq$ its optimal value.

Apoptosis in our model is an all-or-none commitment of individual cells because we assume that MOMP is governed by a bistable switch. As described in ref.[19], the switch is flipped from the "living" state to the "dying" state when [BH3] exceeds [BCL2$_{mit}$] in the mitochondrial outer membrane. The all-or-none nature of the transition in the model is reflected in the fact that most cells survive for $C <$ 5 and most cells die for $C > 6$. For $5 < C < 6$; only a fraction of cells die because of the lognormal distribution we assume for [BCL2$_{mit}$]. Experimental confirmation of the "threshold" effects in **Figure 2d,e** provide strong support for a "bistable switch" underlying MOMP, an assumption of the model that is still a subject of some disagreement among theoreticians.[19,20]

**Quantitative comparison of the model to autophagy and apoptosis in populations of RPT cells**

For quantitative comparison, we chose an "optimal" set of parameter values (**Table 3**) to best fit the data (**Supplementary Table S1**), as detailed in **Supplementary Text S3**. Recognizing that the available data are insufficient to constrain the values of all the parameters in our model, we generated a collection of 3,758 parameter sets that also provided good fits to the data. Simulations using these acceptable parameter sets enable us to compute error bars for our simulation results.

In **Figure 3a**, we plot the model's prediction of autophagy level in response to a cisplatin dose of 20 $\mu$M ($C$ = 20). For each of the acceptable parameter sets, we simulated the response of 100 cells, each with a particular value of [BCL2$_{mit}$]. For each parameter set, we compute, from the sample of 100 cells, the mean value of [ATPHG]($t$) and the percentage of apoptotic cells at each point in time. Then, from the sample of 3,758 acceptable parameter sets, we plot the mean (black line) $\pm$ one standard deviation of autophagy and percent apoptosis as functions of time. The simulations compare very favorably with the experimental data (circles) from figure 6d of Periyasamy-Thandavan et al.[18] The experimental points are singlets and said, by the authors, to be "*representative of at least three separate experiments.*" To get an idea of the variability of these measurements, one should compare figures 1a, 4d, 5d, and 6d in ref.[18]. In these experiments, the level of autophagy increases to a maximum at ~6 hours, after which negative feedback from Ca$^{2+}$ causes the level of autophagy to decrease. As cells become apoptotic, the average level of autophagy drops further. The time course of apoptosis in our simulated population (**Figure 3b**) is in good agreement with the experimental observations in figure 1e of Periyasamy-Thandavan et al.[18]

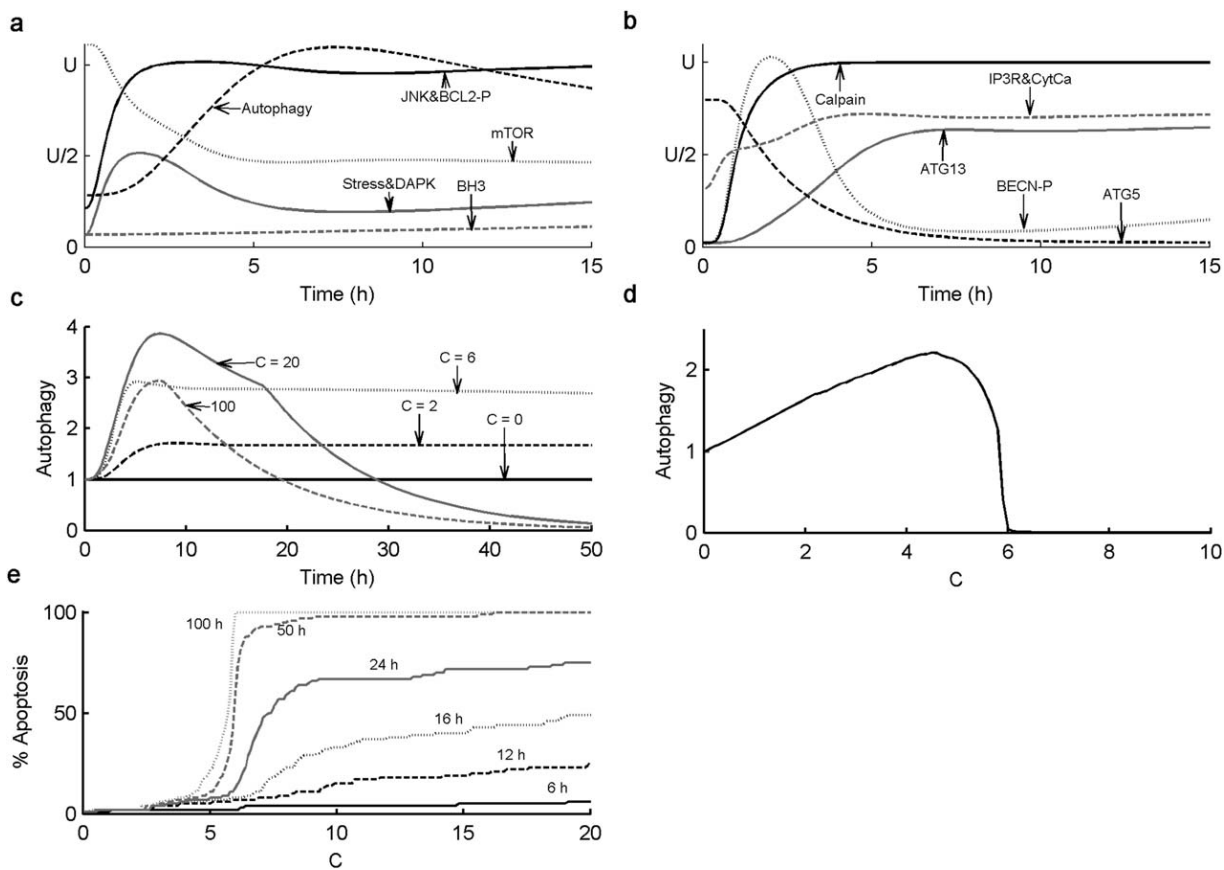

**Figure 2** Qualitative properties of the model. (**a,b**) Simulated time course of the autophagy–apoptosis model for an "average" cell with [BCL2$_{mit}$] = 0.12. The equations in **Table 2** are solved using the optimal parameter values in **Table 3**, given the initial conditions in **Table 1**, with $C$ = 20, for $0 \leq t \leq 15$. Each curve is plotted in terms of an arbitrary "unit" $U$, as follows: $U_{ATG5}$ = 0.9, $U_{ATG13}$ = 0.8, $U_{ATPHG}$ = 1, $U_{BCL2\text{-}P}$ = 2.5, $U_{BECNP}$ = 2.3, $U_{BH3}$ = 1, $U_{Calpain}$ = 1, $U_{CytCa}$ = 1.2, $U_{DAPK}$ = 1.5, $U_{IP3RF}$ = 1.2, $U_{JNK}$ = 1.3, $U_{mTOR}$ = 0.3, $U_{Stress}$ = 13. (**c**) Simulated time course of the relative level of autophagy, [ATPHG]($t$)/[ATPHG](0), in an "average" cell, with [BCL2$_{mit}$] = 0.12, for different levels of stressor, $C$, from 0 to 100. (**d**) Mean relative level of autophagy at $t$ = 100 hours, in a population of 100 cells, with a lognormal distribution of [BCL2$_{mit}$], as a function of increasing stressor, $C$. For low doses of cisplatin ($C$ < 5), the mean level of autophagy increases steadily to counter the effects of cisplatin-induced stress. For 5 < $C$ < 6, autophagy cannot relieve the stress in all 100 cells and some of them commit apoptosis. For $C$ > 6, all cells are apoptotic and [ATPHG] = 0 by $t$ = 100. (**e**) Percentage of apoptotic cells (in a population of 100 cells with lognormally distributed [BCL2$_{mit}$]) at particular timepoints after stimulation, as functions of increasing levels of stressor, $C$.

Experiments by Periyasamy-Thandavan et al.[18] (their figure 5d) show that BCL-2 overexpression not only decreases the basal level of autophagy but also blocks the activation of autophagy after treatment with 20 μM cisplatin (**Figure 3a**, lower curves). To simulate this experiment, we increased BCL-2 expression by 3-fold in both the ER and mitochondria. The reduced autophagic response (in the model) is due to excess BCL-2 protein binding Beclin-1 and preventing initiation of autophagy. At the same time, cell death is also inhibited because the higher BCL-2 level in mitochondria cannot be overwhelmed by the elevated levels of BH3 in response to cisplatin treatment.

### Inhibition of autophagosome docking
The key function of autophagosomes is to engulf damaged cellular material and then fuse with lysosomes, where the collected material is degraded and recycled. Bafilomycin (BAF) is widely used to block the fusion of autophagosomes with lysosomes, resulting in accumulation of autophagosomes in the cytoplasm. **Figure 3c** displays the increased accumulation of autophagosomes in both experiment and simulation after cells were treated with 20 μM cisplatin in combination with 100 nM BAF. To simulate the effect of BAF treatment in the model, both the absorption rate of autophagosomes ($k_{da}$) and (consequently) the rate of stress reduction by autophagy ($k_{ra}$) are reduced 5-fold. It is interesting to note that, when cells are challenged with 20 μM cisplatin + 100 nM BAF, the maximum autophagy level is the same as that observed with 20 μM cisplatin alone, although cells treated with BAF should have much higher stress than cells treated without BAF. In the model, higher stress strongly activates calpain, which cleaves ATG5 and limits the maximum level of autophagy. If calpain is inhibited, cells are observed to die by excess autophagy,[43] and we see a massive increase in autophagy when calpain is inhibited in a simulated treatment with 20 $\mu$M cisplatin + 100 nM BAF (**Supplementary Figure S2**). This role of calpain may have evolved to avert autophagic death in cells subjected to high stress.

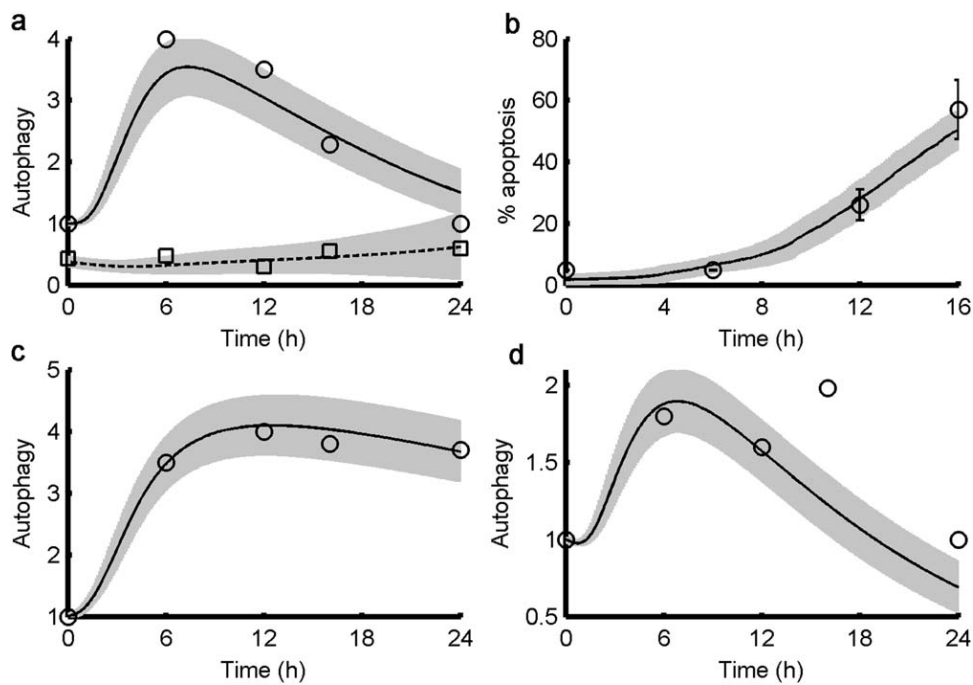

**Figure 3** Time courses of autophagy and apoptosis under cisplatin treatment. In each panel we simulate an experiment from Periyasamy-Thandavan et al.[18] (circles and squares) by solving the equations in **Table 2** using 3,758 different sets of parameter values in the collection of "acceptable" parameter sets, as described in the text and the **Supplementary Material**. The black line plots the mean level of autophagy across all 3,758 simulations, and the gray region spans one standard deviation above and below the mean. (**a**) Time courses of LC3-II and autophagy level for 24-hour treatment with 20 $\mu$M cisplatin alone (circles and solid line) and in conjunction with BCL-2 overexpression (squares and dashed line). The experimental data ("representative of at least three separate experiments") are replotted from Periyasamy-Thandavan et al.[18] (circles from their figure 6d; squares from their figure 5d). In both simulations, $C = 20$; for the case of BCL-2 overexpression, $[BCL2_{ER}] = 9$ and the mean value of $[BCL2_{mit}] = 0.36$ in the simulation. (**b**) Time course of percentage apoptotic cells (circles from figure 1e of Periyasamy-Thandavan et al.[18]). (**c**) Time course of LC3-II (circles from figure 6d of Periyasamy-Thandavan et al.[18]) and simulated autophagy level for 24-hour treatment with 20 $\mu$M cisplatin + 100 nM Bafilomycin (optimal value of $k_{da} = 0.019$ h$^{-1}$ and optimal value of $k_{ra} = 0.77$ h$^{-1}$ in the simulation). (**d**) Time course of LC3-II (circles from figure 6d of Periyasamy-Thandavan et al.[18]) and simulated autophagy level for 24-hour treatment with 20 $\mu$M cisplatin + 10 mM 3-methyladenine (optimal value of $k_a = 0.88$ h$^{-1}$ in the simulation).

Impaired fusion of autophagosomes with lysosomes puts the cell under increased stress and should lead to increased apoptosis. In the experiment, 55% of cells were apoptotic at 12 hours, compared to 53% of the simulated cells (**Figure 4a**). The model predicts that 89% of cells will be apoptotic at 24 hours (**Figure 4b**).

### Inhibition of autophagosome formation
3-MA inhibits formation of autophagosomes. In response to treatment with 20 $\mu$M cisplatin + 10 mM 3-MA, both experiments and simulations show a decreased level of autophagy (**Figure 3d**) compared to treatment with 20 $\mu$M cisplatin alone (**Figure 3a**). To simulate the effect of 3-MA, the parameter controlling formation of autophagosomes ($k_a$) was decreased 2-fold. As with BAF, the decrease in autophagic recycling of cellular material caused by 3-MA should increase cell death. Experimentally, 46% of cells were apoptotic at 12 hours compared to 38% of the simulated cells at this point (**Figure 4a**). The level of apoptosis is not quite as high as for BAF treatment, suggesting higher autophagic recycling in the case of 3-MA treatment. Similarly, inhibition of autophagy, by knocking down Beclin-1 while treating cells with 20 $\mu$M cisplatin, also results in increased apoptosis, both experimentally (55% at 16 hours) and in simulation (57%). (In the simulation, we reduced the initial value of $[BECN1]_T$ by 50%.) As expected, the model predicts significantly higher percentages of apoptotic cells at 24 hours in all cases (**Figure 4b**). We suggest that future experiments measuring apoptosis in cells responding to high cisplatin stress be carried out to 24 hours and beyond, to determine whether most cells eventually commit apoptosis, as predicted by our model (**Figure 3e**).

Our model also accounts for the proapoptotic effect of 3-MA at a cisplatin dose of 5 $\mu$M (**Figure 4a**, as compared to figure 7c of Periyasamy-Thandavan et al.[18]). Because very few cells commit apoptosis at a dose of 5 $\mu$M cisplatin, we cannot associate this dose with $C = 5$, which lies at the cusp of life-or-death in our model. Although the model was parameterized to fit data at a cisplatin dose of 20 μM with $C = 20$, there is no reason to expect that the parameter $C$, which measures metabolic stress, should bear a linear relation to cisplatin dose. Hence, we choose $C = 2.5$, where cells are robustly surviving in the model, to represent the case of cisplatin = 5 μM in experiments.

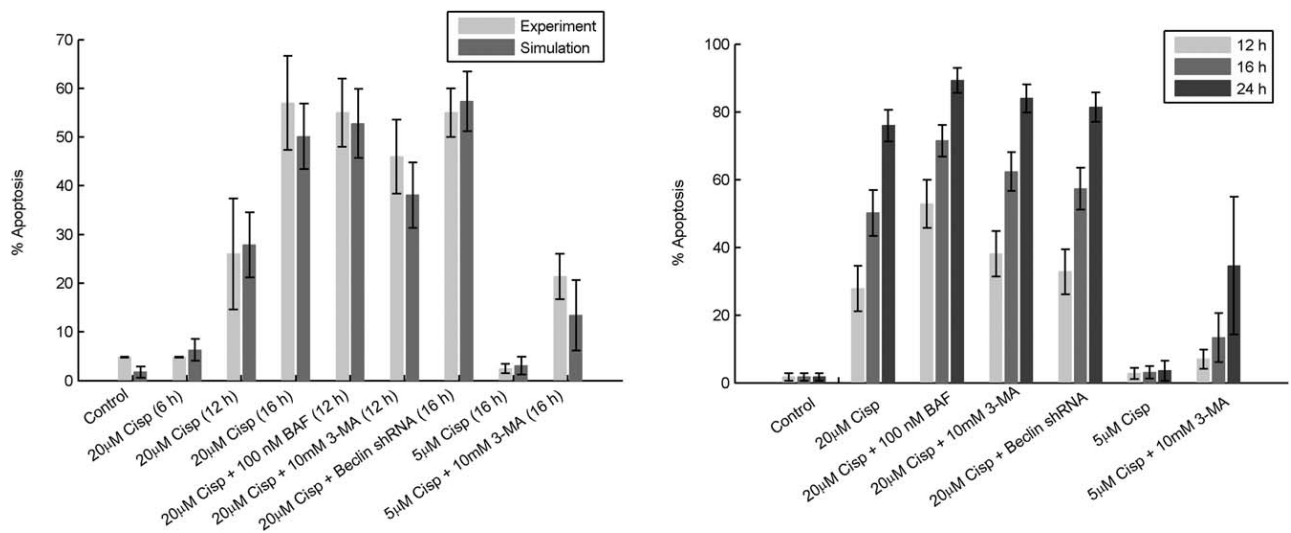

**Figure 4** Percentage of apoptotic cells. (**a**) Percentage of apoptotic cells (mean ± one standard deviation) in experiment (light gray bars) and simulation (medium gray bars) under various treatments at the indicated timepoint. The experimental data are replotted from Periyasamy-Thandavan *et al*.[18] (their figures 7b,c and 8d). "Four fields with ∼200 cells per field were evaluated in each dish to estimate the percentage of cells with typical apoptotic morphology." The simulations are done as described in the legend to **Figure 3**. (**b**) For each of the experimental conditions in panel (a), we plot the percentage of apoptotic cells (mean ± one standard deviation) in simulations at 12 hours (light gray), 16 hours (medium gray), and 24 hours (dark gray). The 24-hour timepoints are predictions of the model.

To see how all variables of an "average" cell respond to all the experimental conditions considered in this section, see **Supplementary Figure S1**.

## DISCUSSION

Recent advances in molecular cell biology indicate that anticancer therapies promote cellular stress, which can trigger both autophagy and apoptosis.[3,44] Hence, understanding the interactions among the molecular regulators of these major cell survival/death pathways is critical to solving clinical issues associated with drug efficacy and side effects. In this study, we show that a mathematical model of cellular stress can capture the prosurvival and prodeath responses of cells in qualitative terms and be fitted accurately to quantitative time-course data of autophagy and apoptosis measured by Periyasamy-Thandavan *et al*.[18] The model predicts the time courses of regulatory proteins (**Figure 2a,b**) and the long-term extent of apoptosis (**Figures 2e** and **4b**), which were not measured in the original experiments. The model predicts that, for the rat kidney cells studied here, the role of cytoplasmic calcium ions in upregulating apoptosis proceeds less through calcineurin's activation of proapoptotic proteins than through calpain's truncation of ATG5. Future quantitative measurements of how cells respond to cytotoxic stress will allow us to test this model and build new versions with more predictive power. In addition, experiments that interfere with apoptosis in cells responding to cisplatin, e.g., by manipulating apoptosis activators and inhibitors, such as second mitochondria-derived activator of caspases (SMAC) and X-linked inhibitor of apoptosis protein (XIAP), will provide data to further support and improve the model.

Cell stress responses, such as autophagy and apoptosis, are central to determining the responses of cancer patients to pharmacological interventions. For example, autophagy is commonly associated with the acquisition of drug resistance by cancer cells and contributes to the poor patient survival rate of many cancers. On the positive side, autophagy plays a cytoprotective role in cisplatin nephrotoxicity. Because autophagy and apoptosis are governed by an intricate dynamic network of interacting proteins, it is imperative to identify and target key components of this network when designing therapeutic regimens for diseases such as cancer.[45–47] An accurate, cancer-specific mathematical model can advance the pharmacodynamic analysis of prospective anticancer agents and help find novel combinations of existing therapies that increase the death of cancer cells using low doses that spare normal cells.

**Acknowledgment.** This work was supported by US National Institutes of Health award U54-CA149147.

**Conflict of Interest.** The authors have no conflicts of interest in the conduct of this research.

**Author Contributions.** I.T., J.P., R.C., W.T.B. and J.J.T. wrote paper, I.T., A.N.S.-H., W.T.B., and J.J.T. Designed Research, I.T. and J.P. analyzed data, I.T., J.P., A.N.S.-H., R.C., W.T.B., and J.J.T. analyzed data.

## Study Highlights

> **WHAT IS THE CURRENT KNOWLEDGE ON THE TOPIC?**
>
> ✓ Autophagy and apoptosis regulate cell fate in response to stress. Experimental studies provide quantitative data on cell survival and death in response to stress. But no computational models have yet attended to the dynamical interplay of autophagy and apoptosis in relation to the experimental data.
>
> **WHAT QUESTIONS DID THIS STUDY ADDRESS?**
>
> ✓ Can a simple mathematical model of the regulation of autophagy and apoptosis effectively account for the observed responses of rat cells to exposure to the cytotoxic drug cisplatin?
>
> **WHAT THIS STUDY ADDS TO OUR KNOWLEDGE**
>
> ✓ Since the model presented here is consistent with the data used for its validation, it provides an initial step towards better understanding of the complex interactions among cell fate pathways.
>
> **HOW THIS MIGHT CHANGE CLINICAL PHARMACOLOGY AND THERAPEUTICS**
>
> ✓ Optimizing therapeutic protocols, especially those involving multiple drugs, will require a detailed understanding of the balance between survival and death in both diseased and normal cells. Dynamic models, such as the autophagy-apoptosis model presented here, can play a key role in understanding and computing optimal protocols.

Supplementary information accompanies this paper on the *CPT: Pharmacometrics & Systems Pharmacology* website (http://www.wileyonlinelibrary.com/psp4)